\documentclass[12pt]{article} 
\title{Finite Dimensional Representations of Quadratic Algebras with
 Three Generators and Applications
 \footnote{Talk given at the VI International Wigner Symposium, 16-22 August 1999, 
 Instabul, Turkey}}
\author{C. Daskaloyannis\thanks{e:mail address: daskalo@auth.gr}\\
                  {\it  Physics Department,}\\             
        {\it Aristotle university of Thessaloniki,}\\
                {\it  54006 Thessaloniki, Greece}
        }

\date{January 2000}
\begin{document}
\maketitle
 
\begin{abstract}
The finite dimensional representations of associative quadratic 
algebras with three generators are investigated by using a 
technique based on 
 the deformed parafermionic oscillator algebra.
One application on the calculation of the eigenvalues of 
 the two-dimensional superintegrable systems is discussed.  
\end{abstract}

\section{Introduction}\label{sec:Alg}
In classical mechanics, integrable system is a system possessing  
more constants of motion in addition to the energy. A 
comprehencive review of the two-dimensional integrable classical 
systems  is given by Hietarinta\cite{Hietarinta87}, where the 
space was assumed to be flat. The case of non flat space is under 
current investigation\cite{Ranada97}.  

 An interesting subset of the totality of
integrable systems is the set of systems, which possess a maximum 
number of integrals, these systems are termed as superintegrable 
ones. The Coulomb and the harmonic oscillator potentials are the 
most familiar classical superintegrable systems, whose their 
quantum counterpart has nice symmetry properties, which are 
described by the $so(N+1)$ and $su(N)$ Lie algebras. 

The Hamiltonian of the classical systems is a quadratic function 
of the momenta. 
 In the case 
of the flat space all the known two  dimensional superintegrable 
systems with quadratic integrals of motion are simultaneously 
separable in more than two orthogonal coordinate 
systems\cite{Fris}. The integrals of motion of a two dimensional 
superintegrable system in flat space close in a classical 
quadratic Poisson algebra\cite{KaMiPogo96}. The study of the 
quadratic Poisson algebras is a matter under investigation related 
to several branches of physics as: the solution of the classical 
Yang - Baxter equation \cite{Sklyanin83},  the two dimensional 
superintegrable systems in flat space or on the sphere   
 \cite{KaMiPogo96}, the statistics \cite{EssRitt94} or  the case
 of "exactly solvable" classical problems \cite{GLZ-1992}.     
 
 The quantization of classical integrable systems turns 
generally  to quantum integrable systems, but sometime one has to 
add correction terms to the integrals of motion or to the 
Hamiltonian, these correction terms seem to be  of  order ${\cal 
O}(\hbar^2)$ \cite{HietGramm89}. The classical Poisson algebra is 
shifted to some quantum polynomial algebra, the same thing is true 
in the case of quadratic Poisson algebra corresponding to the Yang 
- Baxter equation\cite{Sklyanin83}, which is turned to a quantum 
quadratic associative algebra\cite{Sklyanin84}. The same idea was 
discussed in ref.\cite{GLZ-1992}, where the classical problems, 
which are expressed by a quadratic Poisson algebra are mapped to 
quantum ones described by the corresponding quantum operator 
quadratic algebra. The same shift is indeed true for the 
superintegrable systems, where the classical ones correspond to 
the quantum ones and the classical quadratic Poisson algebra is 
mapped to a  quadratic associative 
algebra\cite{GLZ-R1}--\cite{KaMiHaPo99}. 

In this contribution we study the general form of the quadratic 
algebras, which are encountered in the case of the two dimensional 
quantum superintegrable systems, these algebras are called 
$Qu(3)$.  In references  \cite{BoDasKo94,GLZ-R1,GLZ-R2} was 
conjectured that, the energy eigenvalues correspond to finite 
dimensional representations of the latent quadratic algebras. 
Granovkii et al in \cite{GLZ-1992}   studied the representations 
of the quadratic Askey - Wilson algebras $QAW(3)$. Using there the 
proposed ladder representation, the finite dimensional 
representations are calculated and this method was applied to 
several superintegrable systems \cite{GLZ-R1}--\cite{GLZ-R3}.  
Another method\cite{BoDasKo93,BoDasKo94} for calculating the 
finite dimensional representations is the use of the deformed 
oscillator algebra \cite{Das1} and their finite dimensional 
version which are termed as generalized deformed parafermionic 
algebras\cite{Quesne}.

\section{The Qu(3) Algebra}
 Let consider the quadratic associative algebra generated 
by the generators $\left\{ A,\,  B,\,  C\right\}$, 
 which satisfy the commutation relations
\begin{equation}
\begin{array}{l}
\left[ A, B \right] = C\\ \left[ A , C \right] = \alpha A^2 + 
\beta B^2 + \gamma \left\{ A, B \right\} + \delta A + \epsilon B + 
\zeta \\ \left[ B , C \right] = a A^2 + b B ^2 + c\left\{ A, B 
\right\} + d A + e B + z 
\end{array}
\label{eq:prealgebra} 
\end{equation}
After rotating the generators $A$ and $B$, 
 we can always consider the case
$\beta =0$. 

 The Jacobi equality for the commutator induces the relation
$$ \left[ A, \left[ B, C \right] \right] = \left[ B, \left[ A, C 
\right] \right] $$ the following  relations 
 $$ b = - \gamma , \quad c= - \alpha
\quad \mbox{and} \quad e= - \delta $$ 
 must be satisfied, and consequently the
general form of the quadratic algebra (\ref{eq:prealgebra}) can be 
explicitly written as follows: 
\begin{equation}
\left[ A, B \right] = C \label{eq:algebra1} 
\end{equation}
\begin{equation}
\left[ A , C \right] = \alpha A^2  + \gamma \left\{ A, B \right\} 
+ \delta A + \epsilon B + \zeta \label{eq:algebra2} 
\end{equation}
\begin{equation}
\left[ B , C \right] = a A^2 - \gamma B ^2 - \alpha \left\{ A, B 
\right\} + d A -\delta B + z \label{eq:algebra3} 
\end{equation}
The Casimir of this algebra is given by: 
\begin{equation}
\begin{array}{rl}
K =& C^2 - \alpha \left\{ A^2, B \right\} -\gamma \left\{ A, B^2 
\right\} + ( \alpha \gamma - \delta ) \left\{ A, B \right\}+\\ +& 
(\gamma^2 - \epsilon) B^2 + ( \gamma \delta - 2 \zeta ) B +\\ +& 
\frac{2a}{3} A^3 + ( d + \frac{a \gamma}{3} + \alpha^2) A^2 + ( 
\frac{a \epsilon}{3} + \alpha \delta + 2 z ) A 
\end{array}
\label{eq:Casimir} 
\end{equation}
another useful form of the Casimir of the algebra is given by: 
\begin{equation}
\begin{array}{rl}
K=& C^2 +\frac{2 a}{3} A^3 -\frac{ \alpha}{3} \left\{A,A,B\right\} 
- \frac{ \gamma}{3} \left\{A,B,B\right\} +\\ &+ \left( 
{\frac{2{\alpha^2}}{3}} + d + {\frac{2a\gamma}{3}} \right)A^2   + 
\left( -\epsilon + \frac{2\gamma^2}{3} \right) B^2  +\\ 
&+\left(-\delta + \frac{a \gamma}{3}\right)\{ A, B \} + \left( 
\frac{2\alpha\delta}{3} + \frac{a\epsilon}{3} + 
  \frac{d\gamma}{3} + 2z
\right)A+\\ &+ \left( -\frac{\alpha\epsilon}{3} + 
  \frac{2\delta \gamma}{3} - 2\zeta
\right) B + \frac{\gamma z}{3} - \frac{\alpha\zeta}{3} 
\end{array}
\label{eq:Casimir1} 
\end{equation}

This quadratic algebra has many similarities to the Racah algebra 
$QR(3)$, which is a special case of the Askey - Wilson algebra 
$QAW(3)$. The algebra (\ref{eq:algebra1} --  \ref{eq:algebra3}) 
does not coincide with the Racah algebra $QR(3)$, if $a \ne 0$ in 
the relation (\ref{eq:algebra3}). 
 We shall call this algebra $Qu(3)$ algebra.
 Unless this difference between $Qu(3)$ and $QR(3)$ algebra
  a representation theory can be constructed by following the same procedures
   as they were described by Granovskii, Lutzenko and Zhedanov in ref.
\cite{GLZ-1992,GLZ-R1,GLZ-R2}. In this paper we shall give a 
realization of this algebra using the deformed oscillator 
techniques\cite{Das1}. The finite dimensional representations of 
the algebra  $Qu(3)$ will be constructed by constructing a 
realization of the algebra  $Qu(3)$ with 
 the generalized parafermionic algebra introduced by Quesne\cite{Quesne}.

\section{Deformed Parafermionic Algebra}\label{sec:Para}

Let now consider a realization of the algebra $Qu(3)$, by using 
 of the deformed oscillator   technique, i.e. by using
 a deformed
oscillator algebra\cite{Das1} $\left\{ b^\dagger, b, {\cal N} 
\right\}$, which satisfies the 
\begin{equation}
\left[ {\cal N}, b^\dagger \right] = b^\dagger, \quad \left[ {\cal 
N}, b \right] = -b, \quad b^\dagger b = \Phi\left({\cal N}\right), 
\quad b b^\dagger = \Phi \left({\cal N}+1\right) 
\label{eq:DefOsci} 
\end{equation}
where the function $\Phi(x)$ is a "well behaved" real function 
which satisfies the the the boundary condition: 
\begin{equation}
\Phi(0)=0, \quad  \Phi(x),  \quad \mbox{for} \quad x>0 
 \label{eq:restriction1}
\end{equation}
 As it is well known\cite{Das1} this
constraint imposes the existence a Fock type representation of the 
deformed oscillator algebra, which is bounded by bellow, i.e. 
there is a Fock basis $|n>,\; n=0,1,\ldots$ such that 
\begin{equation}
\begin{array}{l}
{\cal N}|n>=n|n>\\ b^\dagger | n> = \sqrt{ \Phi \left(n+1\right)} 
| n+1>,\quad  n=0,1,\ldots\\ b|0>=0\\ b|n>= \sqrt{ \Phi 
\left(n\right)} | n-1>,\quad  n=1,2,\ldots 
\end{array}
\label{eq:Fock} 
\end{equation}
The Fock representation (\ref{eq:Fock}) is bounded by bellow. 

In the case of nilpotent deformed oscillator algebras, there is a 
positive integer $p$, such that $$ b^{p+1}=0, \quad 
\left(b^\dagger\right)^{p+1}=0 $$ the above equations imply that 
\begin{equation}
\Phi(p+1)=0, 
 \label{eq:restriction2}
\end{equation}
In that case the deformed oscillator (\ref{eq:DefOsci}) has a 
finite dimensional representation, with dimension equal to $p+1$, 
this kind of oscillators are called deformed parafermion 
oscillators of order $p$. 

An interesting property of the deformed parafermionic algebra is 
that the existence of a faithfull finite dimensional 
representation of the algebra implies that: 
\begin{equation}
 {\cal N} \left(  {\cal N} -1 \right) \left(  {\cal N} -2 \right)\cdots \left(  {\cal N} -p \right) = 0
\label{eq:Nrestriction} 
\end{equation}
The above restriction and the constraints (\ref {eq:restriction1}) 
and (\ref {eq:restriction2}) imply that the general form of the 
structure function $\Phi( {\cal N} )$ has the general 
form\cite{Quesne}: $$ \Phi( {\cal N} ) = {\cal N}( p+1-{\cal N}) ( 
a_0 + a_1 {\cal N}+ a_2 {\cal N}^2 +\cdots a_{p-1} {\cal N}^{p-1} 
) $$

\section{Oscillator realization of the algebra $Qu(3)$}

We shall show, that there is a realization of the algebra $Qu(3)$, 
such that 
\begin{eqnarray}
A= A\left({\cal N}\right)\label{eq:A}\\ B=b\left({\cal N}\right)+ 
b^\dagger \rho \left({\cal N}\right)+ \rho \left({\cal N}\right)b 
\label{eq:B} 
\end{eqnarray}
where the $A[x],\; b[x]$ and $\rho(x)$ are functions, which will 
be determined. In that case (\ref{eq:algebra1}) implies: 
\begin{equation}
C=\left[A,B\right]\; \Rightarrow\; C=b^\dagger \Delta A\left({\cal 
N}\right)\rho \left({\cal N}\right) -\rho \left({\cal 
N}\right)\Delta A\left({\cal N}\right) b \label{eq:C} 
\end{equation}
where 
 $$ \Delta A\left({\cal N}\right) = A\left({\cal N}+1\right) 
- A\left({\cal N}\right) $$ 
 Using equations (\ref{eq:A}),  
(\ref{eq:B}) and (\ref{eq:algebra2}) we find: 
\begin{equation}
\begin{array}{rl}
[A,C]=& [ A\left({\cal N}\right), 
 b^\dagger \Delta A\left({\cal N}\right)\rho\left({\cal N}\right)-
\rho\left({\cal N}\right)\Delta A\left({\cal N}\right) b ]=\\ 
 =& b^\dagger \left( \Delta A\left({\cal N}\right)\right)^2 \rho\left({\cal N}\right)+
 \rho\left({\cal N}\right)\left( \Delta A\left({\cal N}\right)\right)^2 b=\\
 =&
 \alpha A^2  + \gamma \left\{ A, B \right\} + \delta A + \epsilon B + \zeta=\\
=& b^\dagger \left( \gamma \left(  A\left({\cal N}+1\right)+ 
A\left({\cal N}\right)\right)+\epsilon \right)\rho\left({\cal 
N}\right)+ 
\\
&+ \rho\left({\cal N}\right) \left( \gamma \left(  A\left({\cal 
N}+1\right)+ A\left({\cal N}\right)\right)+ \epsilon \right)b + \\ 
&+ 
 \alpha A\left({\cal N}\right)^2  + 2\gamma  A\left({\cal N}\right) b\left({\cal N}\right)
  + \delta A\left({\cal N}\right) + \epsilon B\left({\cal N}\right) + \zeta
\end{array}
\label{eq:detail1} 
\end{equation}
therefore we have the following relations: 
\begin{eqnarray}
 \left( \Delta   A\left({\cal N}\right) \right)^2=
 \gamma \left(  A\left({\cal N}+1\right)+ A\left({\cal N}\right)\right)+
 \epsilon
 \label{eq:eqn1}\\
 \alpha A\left({\cal N}\right)^2  + 2\gamma  A\left({\cal N}\right) b\left({\cal N}\right)
  + \delta A\left({\cal N}\right) + \epsilon B\left({\cal N}\right) + \zeta =0
  \label{eq:eqn2}
\end{eqnarray}
while the function $\rho\left({\cal N}\right)$ can be arbitrarily 
determined. In fact  this function can be fixed, in order to have 
a polynomial structure function $\Phi(x)$ for the deformed 
oscillator algebra (\ref{eq:DefOsci}). 

The solutions of equation (\ref{eq:eqn1})  depend on the value of 
the parameter $\gamma$, while the function $b({\cal N})$ is 
uniquely determined by equation (\ref{eq:eqn2}) (provided that 
almost one among the parameters $\gamma$ or $\epsilon$ is not 
zero). 
  At this stage,
 the cases $\gamma \ne 0$ or  $\gamma =0$, should be treated separately.
 We can see that:
\begin{itemize}
\item[{\bf Case 1}:]  $\gamma \ne 0$\\
In that case the solutions of equations (\ref{eq:eqn1}) and 
(\ref{eq:eqn2}) are given by: 
\begin{equation}
 A\left({\cal N}\right) =
 \frac{\gamma}{2} \left(
 ({\cal N}+u)^2-1/4
-\frac{\epsilon}{ \gamma^2} \right) \label{eq:sol1} 
\end{equation}
\begin{equation}
\begin{array}{rl}
b \left({\cal N}\right)=& -{\frac{ \alpha  \left( ({\cal 
N}+u)^2-1/4 \right) }{4}} + {\frac{\alpha\,\epsilon - 
     \delta\,\gamma}{2\,
     {{\gamma}^2}}}
-
\\
&-{\frac{ \alpha\,{{\epsilon}^2} 
           - 2\,\delta\,\epsilon\,
      \gamma + 4\,{{\gamma}^2}\,
      \zeta}{4\,{{\gamma}^4}}}
\frac{1}{  \left( ({\cal N}+u)^2-1/4 \right) } 
\end{array}
\label{eq:sol2} 
\end{equation}

\item[{\bf Case 2}:]  $\gamma = 0, \; \epsilon \ne 0 $\\
 The solutions of equations (\ref{eq:eqn1}) and (\ref{eq:eqn2}) are given by:
\begin{equation}
A({\cal N}) = \sqrt{\epsilon}  \left( {\cal N}+u \right) 
\label{eq:sol1a} 
\end{equation}
\begin{equation}
b({\cal N}) =-\alpha \left( {\cal N}+u \right)^2- \frac{\delta}{ 
\sqrt{\epsilon} }   \left( {\cal N}+u \right) 
  -
\frac{\zeta}{\epsilon} \label{eq:sol2a} 
\end{equation}
\end{itemize}

The constant $u$ will be determined later. 

Using the above definitions of equations $A({\cal N})$ and 
$b({\cal N})$, the left hand side and right hand side of equation 
(\ref{eq:algebra3}) gives the following equation: 
\begin{equation}
\begin{array}{l}
2\,\Phi({\cal N}+1)\left( \Delta A\left({\cal N}\right) 
+\frac{\gamma}{2} \right)  \rho({\cal N}) 
-
2\,\Phi({\cal N})\left( \Delta A\left({\cal N}-1\right) 
-\frac{\gamma}{2} \right) \rho({\cal N}-1) =\\ =a A^2\left({\cal 
N}\right) -\gamma b^2({\cal N}) -2 \alpha A\left({\cal N}\right) 
b({\cal N}) +d A\left({\cal N}\right)-\delta b({\cal N})+z 
\end{array}
\label{eq:basic1} 
\end{equation}
Equation (\ref{eq:Casimir}) gives the following relation: 
\begin{equation}
\begin{array}{rl}
K=\\ =& \Phi({\cal N}+1)\left( \gamma^2 - \epsilon - 2 \gamma 
A\left({\cal N}\right)  - \Delta A^2\left({\cal N}\right) \right) 
\rho({\cal N}) 
 +\\
&+ \Phi({\cal N})\left( \gamma^2 - \epsilon - 2 \gamma 
A\left({\cal N}\right)  - \Delta A^2\left({\cal N}-1\right) 
\right) \rho({\cal N}-1) -\\ &-2 \alpha A^2\left({\cal N}\right) 
b({\cal N}) +\left( \gamma^2 - \epsilon - 2 \gamma  A\left({\cal 
N}\right) \right) b^2({\cal N})+\\ &+ 2 \left( \alpha \gamma - 
\delta \right)  A\left({\cal N}\right) b({\cal N}) +\left( \gamma 
\delta - 2\zeta \right) b({\cal N})+ 
\\
&+ \frac{2}{3}a A^3\left({\cal N}\right) + \left( d + \frac{1}{3} 
a\gamma + \alpha^2 \right)A^2\left({\cal N}\right)+\\ &+\left( 
\frac{1}{3} a \epsilon + \alpha \delta +2 z \right)A\left({\cal 
N}\right) 
\end{array}
\label{eq:basic2} 
\end{equation}

Equations (\ref{eq:basic1}) and (\ref{eq:basic2}) are linear 
functions of the expressions $\Phi\left({\cal N}\right)$ and 
$\Phi\left({\cal N}+1\right)$, then 
 the function
$\Phi\left({\cal N}\right)$ can be determined, if the function $ 
\rho({\cal N})$ is given. The solution of this system, i.e. the 
function $\Phi\left({\cal N}\right)$ depends on two parameters $u$ 
and $K$ and it is given by the following formulae: 
\begin{itemize}
\item[{\bf Case 1}:]  $\gamma \ne 0$\\
$$ \rho( {\cal N} ) = \frac{1}{3\cdot 2^{12}\cdot \gamma^8 ( {\cal 
N}+u) ( 1 + {\cal N}+u ) ( 1 +2 ( {\cal N}+u) )^2} $$ and 
\begin{equation}
\begin{array}{l}
\Phi({\cal N}) =-3072 \gamma^6 K (-1 + 2 ({\cal N}+u))^2-\\ - 48 
\gamma^6 
   (\alpha^2  \epsilon - \alpha  \delta  \gamma + a \epsilon  \gamma - d  \gamma^2) \cdot\\
\cdot (-3 + 2  ({\cal N}+u))  (-1 + 2  ({\cal N}+u))^4 
   (1 + 2  ({\cal N}+u)) + \\
+\gamma^8  (3  \alpha^2 + 4  a  \gamma)  (-3 + 2  ({\cal N}+u))^2  
(-1 + 2  ({\cal N}+u))^4 
   (1 + 2  ({\cal N}+u))^2 +\\
+ 768  (\alpha  \epsilon^2 - 2  \delta  \epsilon  \gamma + 4  
\gamma^2  \zeta)^2 +\\ + 
  32 \gamma^4 (-1 + 2 ({\cal N}+u))^2 (-1 - 12 ({\cal N}+u) + 12 ({\cal N}+u)^2) \cdot \\
  \cdot  (3 \alpha^2 \epsilon^2 - 6 \alpha \delta \epsilon \gamma + 2 a \epsilon^2 \gamma + 2 \delta^2 \gamma^2 -
     4 d \epsilon \gamma^2 + 8 \gamma^3 z + 4 \alpha \gamma^2 \zeta) -\\
-
  256 \gamma^2(-1 + 2 ({\cal N}+u))^2 \cdot\\
\cdot (3 \alpha^2 \epsilon^3 - 9 \alpha \delta \epsilon^2 \gamma + 
     a \epsilon^3 \gamma + 6 \delta^2 \epsilon \gamma^2 - 3 d \epsilon^2 \gamma^2 + 2 \delta^2 \gamma^4 + \\
+ 
     2 d \epsilon \gamma^4 + 12 \epsilon \gamma^3 z
- 4 \gamma^5 z + 12 \alpha \epsilon \gamma^2 \zeta - 
     12 \delta \gamma^3 \zeta + 4 \alpha \gamma^4 \zeta)
 \end{array}
\label{eq:Phi1} 
\end{equation}
\item[{\bf Case 2}:]  $\gamma = 0, \; \epsilon \ne 0 $\\
$$\rho({\cal N}) =1$$ 
\begin{equation}
\begin{array}{l}
\Phi( {\cal N} )=\\ = \frac{1}{4} 
 \left( -\frac{K}{\epsilon} - \frac{z}{\sqrt{\epsilon}} - \frac{\delta}{\sqrt{\epsilon}}   \frac{\zeta}{\epsilon} +
\frac{\zeta^2}{\epsilon^2} \right) -\\ -\frac{1}{12}\Big( 3 d  - a 
\sqrt{\epsilon} - 3 \alpha  \frac{\delta}{\sqrt{\epsilon}} + 3 
\left(\frac{\delta}{\sqrt{\epsilon}}\right)^2- 6  
\frac{z}{\sqrt{\epsilon}} +6 \alpha \frac{\zeta}{\epsilon}- 6  
\frac{\delta}{\sqrt{\epsilon}}   \frac{\zeta}{\epsilon}\Big)  
({\cal N}+u)\\ + \frac{1}{4} \left( \alpha^2 + d - a 
\sqrt{\epsilon} - 3 \alpha \frac{\delta}{\sqrt{\epsilon}}+ \left( 
\frac{\delta}{\sqrt{\epsilon}}\right)^2+ 2 \alpha 
\frac{\zeta}{\epsilon} \right)  ({\cal N}+u)^2-\\ -\frac{1}{6} 
\left( 3 \alpha^2 - a \sqrt{\epsilon} - 3 \alpha 
\frac{\delta}{\sqrt{\epsilon}} \right) ({\cal N}+u)^3 +\frac{1}{4}  
\alpha^2 ({\cal N}+u)^4 
\end{array}
\label{eq:Phi2} 
\end{equation}
\end{itemize}

\section{Finite dimensional representations of the algebra $Qu(3)$}
Let consider a representation of the algebra $Qu(3)$, which is 
diagonal to the generator $A$ and the Casimir $K$. Using the 
parafermionic realization defined by equations (\ref{eq:A}) and 
(\ref{eq:B}), we see that this a representation diagonal to the 
parafermionic number operator ${\cal N}$ and the  Casimir $K$. The 
basis of a such representation corresponds to the Fock basis of 
the parafermionic oscillator, i.e. the vectors $|k, \, n >,\; 
n=0,1,\ldots $of the carrier Fock space satisfy the equations $$ 
{\cal N}   |k, \, n >= n|k, \, n >, \quad K|k, \, n >= k |k, \, n 
> $$ The structure function (\ref{eq:Phi1}) (or respectively 
(\ref{eq:Phi1}) ) depend on the eigenvalues of the of the 
parafermionic number operator ${\cal N}$ and the  Casimir $K$. The 
vectors $|k, \, n >$ are also eigenvectors of the generator $A$, 
i.e. $$ A  |k, \, n >= A(k,n)|k, \, n > $$ 

In the case $\gamma \ne 0$ we find from equation (\ref{eq:sol1}) 
$$ A\left(k,n\right) = 
 \frac{\gamma}{2} \left(
 (n+u)^2-1/4
-\frac{\epsilon}{ \gamma^2} \right) $$ 

In the case $\gamma = 0,\; \epsilon \ne 0$ we find from equation 
(\ref{eq:sol1a}) $$ A(k,n) = \sqrt{\epsilon}  \left( n+u \right) 
$$ then the parameter $u=u(k,p)$ is a solution of the system of 
equations (\ref{eq:system}).

If the deformed oscillator corresponds to a deformed Parafermionic 
oscillator of order $p$ then the two parameters of the calculation  
$k$  and $u$ should satisfy the constrints (\ref{eq:restriction1})  
and (\ref{eq:restriction2}) the system: 
\begin{equation}
\begin{array}{c}
\Phi(0,u,k)=0 \\ \Phi(p+1,u,k)=0 
\end{array}
\label{eq:system} 
\end{equation}
then the parameter $u=u(k,p)$ is a solution of the system of 
equations (\ref{eq:system}). 

Generally there are many solutions of the above system, but a 
unitary representation of the deformed parafermionic oscillator is 
restrained by the additional restriction $$ \Phi(x) >0, \quad 
\mbox{for}  \quad 0<x<p+1 $$ We must point out that the system 
(\ref{eq:system}) corresponds to a  representation with dimension 
equal to $p+1$. 

\section{Application of the case $\gamma=0$}

In this section, we shall give an example of the calculation of 
eigenvalues of a superintegrable two-dimensional system, by using 
the methods of the previous section. The  calculation by an 
empirical method was performed in \cite{BoDasKo94} and the 
solution of the same problem by using separation of variables was 
studied in \cite{KaMiPogo96}.  Here in order to show the effects 
of the quantization procedure we don't use $\hbar=1$ as it was 
considered in references  \cite{BoDasKo94} and  \cite{KaMiPogo96}. 
That means that the following commutation relations are taken in 
consideration: 
\[ 
[x,p_x]= i \hbar, \qquad [x,p_x]= i \hbar 
\] 

 The  superintegrable Holt  
system corresponds to the Hamiltonian: 
\begin{equation}
H=\frac{1}{2}\left( p_x^2 + p_y^2 \right) 
+\frac{\omega^2}{2}\left(4 x^2 + y^2\right) + 
\frac{k^2-\frac{1}{4}}{y^2}. 
 \label{eq:HoltHamiltonian} 
\end{equation}
This superintegrable system with two  integrals: $$A= p_x^2+4 
\omega^2 x^2, \quad 
 \mbox{and}
  \quad
   B=
 \left\{p_y, xp_y-yp_x\right\}
    -2\omega^2 x y^2
        -2(1/4-k^2) \frac{x}{y^2}  
$$ 
 From the above definitions we can verify that:
$$ \left[{H},{A}\right]=0, \quad 
 \left[{H},{B}\right]=0,  $$
and $$ \left[{A},{B}\right]=C, \quad 
 \left[{A},{C}\right]= 16 \hbar^2  \omega^2 B,
$$ 
 $$ 
 \left[{B},{C}\right] = 24 h^2 A^2-64 \hbar^2 H A+
  8 \hbar^2 \left(
 4 H^2   +\omega^2(1 - 4 k^2+ 3 \hbar^2 )\right)
 $$ 
 The above algebra is a quadratic algebra $Qu(3)$ of the form 
(\ref{eq:algebra1}--\ref{eq:algebra3}), corresponding to the 
following values of the coefficients: 
 $$ 
\begin{array}{c}
\alpha=0, \quad \gamma=0, \quad \delta=0, \quad \epsilon=16 
\hbar^2  \omega^2, \quad \zeta=0 \\ 
 a=24 \hbar^2,\quad d=-64 \hbar^2 H,\quad
  z= 8 \hbar^2 \left(
 4 H^2   +\omega^2(1 - 4 k^2+ 3 \hbar^2 )\right)
\end{array}
$$ 
 The value of the Casimir operator (\ref{eq:Casimir}) is given 
by: 
 $$
 \begin{array}{rl}
  K=&C^2 +\\
    &+16 \hbar^2 A^3 -64 \hbar^2 H A^2    
    -16 \hbar^2\omega^2 B^2
    +\\
    &+16 \hbar^2 \left(
 4 H^2   +\omega^2(1 - 4 k^2+ 11 \hbar^2 )\right) A=\\ 
=& 256 \hbar^4 \omega^2 H 
\end{array}
 $$ 

The representation of the above algebra, which is diagonal to the 
Hamiltonian $H$ and 
 to the integral of motion $A$, corresponds to the eigenvalues of the energy $E$
  and the eigenvalue of the Casimir equal to $256 \hbar^4 \omega^2 E$
In that case, equations (\ref{eq:basic1}) and (\ref{eq:basic2}), 
 which determine the function $\Phi(x)$, are respectively:
$$ 
\begin{array}{l}
-32 \hbar^2  E^2 - 8 \hbar\omega{\Phi(x)} + 
8\hbar\omega{\Phi(x+1)} - 8\hbar^2\omega^2 -\\ 
    -24\hbar^4\omega^2 + 32\hbar^2 k^2\omega^2 + 256\hbar^3\ E\omega (x+u) - 
    384\hbar^4\omega^2 (x+u)^2=0
\end{array}
$$ 

$$ 
\begin{array}{l}
 -32\,\left( {\Phi(x)} + {\Phi(x+1)} 
     \right) \,{\hbar^2}\,{\omega^2} +\\ 
  +64\,{\hbar^3}\,\omega\,\left( 4\,{E^2} + {\omega^2} + 
     11\,{\hbar^2}\,{\omega^2} - 4\,{k^2}\,{\omega^2} \right) \,
   (x+u) -\\
   - 1024\,{\hbar^4}\,E\,{\omega^2}\,{(x+u)^2} + 
  1024\,{\hbar^5}\,{\omega^3}\,{(x+u)^3}=0
\end{array}
$$ 

The above equations can be solved and we find that: $$ 
\begin{array}{rl}
\Phi(x)=&\displaystyle {\frac{-\ \hbar\,\left( 4\,{E^2} + 
8\,\hbar\,E\,\omega + 
          {\omega^2} + 3\,{\hbar^2}\,{\omega^2} - 
          4\,{k^2}\,{\omega^2} \right)   }{2\,\omega}
    }\\ &\displaystyle+ {\frac{\hbar\,\left( 4\,{E^2} + 16\,\hbar\,E\,\omega + 
        {\omega^2} + 11\,{\hbar^2}\,{\omega^2} - 4\,{k^2}\,{\omega^2}
         \right) \,(x+u)}{\omega}} -\\
         &\displaystyle- 
  8\,{\hbar^2}\,\left( 2\,E + 3\,\hbar\,\omega \right) \,
   {(x+u)^2} + 16\,{\hbar^3}\,\omega\,{(x+u)^3}
\end{array}
$$ 

 The two parameters of this equation are the parameter $u$ 
and the eigenvalue $E$ of the energy $H$, therefore we can solve 
the system: 
 $$ \Phi(0)=0, \quad \Phi(p+1)=0 $$ 
 and we find two 
solutions: 
 $$ u=\frac{1}{2}, \quad
  E=\frac{\omega }{2}\left(
  4\,\hbar\,( 1 + p )
  \pm   {\sqrt{-1 + {\hbar^2} + 4\,{k^2}}}\,     
     \right)
   $$

\section{Application of the case $\gamma \ne 0$}

Let consider the case of a potential on the two dimensional 
hyperboloid taken from ref \cite{KaMiHaPo99} (case of the 
potential $V_1$). 

The two dimensional hyperpoloid is characterized by the cartesian 
coordinates $\omega_0,\omega_1,\omega_2$, which obey to the 
restriction $\omega_0^2-( \omega_1^2+ \omega_1^2) =1$. The 
Hamiltonian is given by 
\[ 
H= - \frac{1}{2} \triangle_{LB} +V 
\]   
where $\triangle_{LB}$ is the Laplace - Beltrami operator for the 
details see ref  \cite{KaMiHaPo99}, where 
\[ 
V=\frac{\alpha^2}{\omega_2^2}- \frac{\gamma^2}{ ( 
\omega_0-\omega_1)^2} + \beta^2 \frac{\omega_0+\omega_1}{ 
\omega_0-\omega_1)^3} 
\]
The two integrals are 
\[
A= \left( \omega_0 \partial_{\omega_1} +  \omega_1 
\partial_{\omega_0}\right)^2 - 
2 \beta^2 \left( \frac{\omega_0+\omega_1}{ \omega_0-\omega_1 } 
\right)^2+ 2 \gamma^2  \frac{\omega_0+\omega_1}{ \omega_0-\omega_1 
} 
\]
and 
\[
B= \left(\omega _0\partial _{\omega _2}+\omega _2\partial _{\omega 
_0}- \omega_1\partial_{\omega_2} + \omega_2 
\partial_{\omega _1}\right)^2-\frac{2\beta^2 
\omega_2^2}{(\omega_0-\omega_1)^2} -\frac{2\alpha^2 
(\omega_0-\omega_1)^2}{\omega_2^2} 
\]
The operators $H,\;A$ and $B$ satisfy the following commutation 
relations: 
\[ 
\left[ H, A \right]= 0, \qquad \left[ H, A \right]= 0 
\]
\[
[ A, C] = 8 \{ A, B\} +16\gamma^2 A- 16B  - 
 16 \gamma^2(1-2\alpha^2 -2 H ) 
\]
\[
[ B, C] = -8 B^2  -32 \beta^2 A - 16 \gamma^2 B - 16\beta^2(1 - 
4\alpha^2 + 4 H) 
\]
 The above algebra is a quadratic algebra $Qu(3)$ of the form 
(\ref{eq:algebra1}--\ref{eq:algebra3}), corresponding to the 
following values of the coefficients: 
 $$ 
\begin{array}{c}
\alpha=0, \quad \gamma=8, \quad \delta=16\gamma^2, \quad \epsilon= 
- 16 , \quad \zeta= - 
 16 \gamma^2(1-2\alpha^2 -2 H ) \\ 
 a=0,\quad d=-32 \beta^2,\quad
  z= - 16\beta^2(1 - 4\alpha^2 + 4 H) 
\end{array}
$$ The value of the Casimir operator (\ref{eq:Casimir}) is given 
by: 
 $$
 \begin{array}{rl}
  K=&C^2 -\\
   &-\frac{8}{3}\{A, B, B\} + \frac{176}{3}B^2 - 32\beta^2 A^2
- 16\gamma^2 \{A, B\} -\\ &-  \left(64\alpha^2\gamma^2 + 
64\gamma^2 H- \frac{352}{3}\gamma^2\right) B - \left(\frac{352}{3} 
-128 \alpha^2+128\beta^2 H\right) \beta^2 A -  
\\
&-\frac{128}{3}\beta^2(1 - 4\alpha^2 + 4 H) =\\ =& 16\,( 
-4\,\beta^2 + 
    8\,\alpha^2\,\beta^2 + 
    8\,{{\alpha}^4}\,\beta^2 - 
    3\,{{\gamma^4}} + \\
    &+
    8\,\alpha^2\,{{\gamma^4}} - 
    8\,\beta^2\,H + 
    16\,\alpha^2\,\beta^2\,H + 
    8\,\beta^2\,{H^2} )
\end{array}
 $$ 
The representation of the above algebra, which is diagonal to the 
Hamiltonian $H$, 
 to the integral of motion $A$,  corresponds to the eigenvalues of the energy $E$
  and the eigenvalue of the Casimir equal to above cited value.

In that case, equations (\ref{eq:basic1}) and (\ref{eq:basic2}), 
 which determine the function $\Phi(x)$, give the form 
 (\ref{eq:Phi1}):
\[
 \begin{array}{rl}
 \Phi(x)=& 
-3\cdot 2^{34} \left( 2\,\beta^2 - {{\gamma^4}} - 
  8\,\beta^2\,\left( u + x \right)  + 
  8\,\beta^2\,{{\left( u + x \right) }^2}
\right) \cdot \\ & 
 \cdot ( -\alpha^2 + {{\alpha^4}} + E + 
  2  \alpha^2  E + {E^2} + 
  \left( -1 + 4  \alpha^2 - 4  E \right)   
   \left( u + x \right)  +\\
   &+ 
  \left( 5 - 4  \alpha^2 + 4  E\right)   
   {{\left( u + x \right) }^2} - 
  8  {{\left( u + x \right) }^3} + 
  4   {{\left( u + x \right) }^4}
) 
\end{array}
\]
The two parameters of this equation are the parameter $u$ and the 
eigenvalue $E$ of the energy $H$, therefore we can solve the 
system: 
 $$ \Phi(0)=0, \quad \Phi(p+1)=0 $$ 
 and we find the 
solution: 
\[
u=\frac{1}{2}\left(1- \frac{\gamma^2}{\sqrt{2}\beta}\right), \quad 
 E=-\frac{1}{2}\left(2p+2+\sqrt{2\alpha^2+1/4}
-\frac{\gamma^2}{\sqrt{2}\beta}\right)^2 +\frac{1}{8} 
\]

\section{Discussion}
From the above discussion, we have shown how to calculate finite 
dimensional representations of the $Qu(3)$ algebra and we have 
given an application of this method in the calculation of the 
energy eigenvalues of the superintegrable systems. The systematic 
algebraic study of all the known superintegrable systems is under 
investigation.

\end{document}